\documentclass[journal,transmag]{IEEEtran}

\usepackage{hyperref}
\usepackage{amsmath}
\usepackage{amsthm}

\newtheorem{lemma}{Lemma}
\newcommand{\killpunct}[1]{}

\ifCLASSINFOpdf
\usepackage[pdftex]{graphicx}

\else

\fi

\begin{document}

\title{An Architecture for In-Vehicle Networks}

\author{\IEEEauthorblockN{Jean Walrand\IEEEauthorrefmark{1}~\IEEEmembership{Life Fellow,~IEEE},
Max Turner\IEEEauthorrefmark{2}~\IEEEmembership{Member,~IEEE}, and
Roy Myers\IEEEauthorrefmark{2}~\IEEEmembership{Member,~IEEE}}
\IEEEauthorblockA{\IEEEauthorrefmark{1}Department of Electrical and Computer Sciences,
University of California, Berkeley, CA 94720 USA}
\IEEEauthorblockA{\IEEEauthorrefmark{2}Ethernovia, San Jose, CA, USA}
\thanks{This work has been submitted to the IEEE for possible publication. Copyright may be transferred without notice, after which this version may no longer be accessible.
Corresponding author: J. Walrand (email: walrand@berkeley.edu).}}

\markboth{}%
{Walrand \MakeLowercase{\textit{et al.}}: An Architecture for In-Vehicle Networks}

\IEEEtitleabstractindextext{%
\begin{abstract}
As vehicles get equipped with increasingly complex sensors and processors, the communication requirements
become more demanding. Traditionally, vehicles have used specialized networking technologies designed
to guarantee bounded latencies, such a the Controller Area Network (CAN) bus.  Recently, some have used
dedicated technologies to transport signals from cameras, lidars, radars, and ultrasonic sensors.  
In parallel, IEEE working groups are defining Ethernet standards for time-sensitive networks (TSN).  This
paper describes an Ethernet-based architecture with provable guaranteed performance and simple
configuration that is suitable for supporting the communication requirements of many vehicles.
\end{abstract}

}

\maketitle

\IEEEdisplaynontitleabstractindextext

\IEEEpeerreviewmaketitle

\section{Introduction}

Since the early 1990s, automobile manufacturers have adopted the CAN bus technology to connect
microcontrollers and devices \cite{CANinW140}. While the automotive industry has developed and used multiple other communication technologies 
like MOST \cite{MOSTbook} and FlexRay \cite{FlexRayX5} over the years, CAN is still to be found in almost every car manufactured worldwide today.
In 2008 the first vehicle using Ethernet was put into production \cite{AutoEth12Years}. The main motivation at the time was faster software updates which started to consume several hours over the $500$kbps diagnostic CAN connection.
The first cameras used on vehicles to help the driver with maneuvering during parking had analog connections as these would fit in well with other video sources like television.

Today many modern automobiles are equipped with cameras, radars, ultrasonic detectors, and other sensors to
support various degrees of driver assistance or self-driving
capabilities. To connect those sensors to processors, actuators, and
user interfaces, these automobiles often use a combination of network technologies (see \cite{AutoEthBook} and \cite{c2b}). 
The addition of these modern sensors further adds to the cost, weight, and complexity of the already
cumbersome wiring harness of the automobile.  To simplify this wiring, the industry is exploring the use
of Ethernet technology.  This technology has a number of important characteristics: it is standardized, has
proved its flexibility over four decades, is high-speed, and it can integrate the transport of many different types
of signals into one unified network. 
Time triggered bus systems like FlexRay and MOST had been developed specifically to address use-cases in electro-mechanical control loops and audio distribution respectively.
One immediate concern when considering Ethernet as a transport technology inside a vehicle is that, in its
basic form, it does not provide guarantees on communication delays.  However, the development by IEEE of
a family of standards for time-sensitive networks is addressing that issue \cite{tsn} (see \cite{cisco} for an overview).  These standards combine
various traffic policing, shaping (see e.g., \cite{IEEE-ATS}-\cite{Ubits xtraffic}), and scheduling
(see e.g., \cite{IEEE-CQF}) to provide precise guarantees on the latency of packets
across the network.

Unfortunately, it is not easy to select the combination of policiers, shapers, and schedulers,
and their parameters
to achieve specific latency and throughput objectives.  See \cite{Queck}-\cite{Thangamuthu15} for
the analysis of some subsets of the IEEE protocols. 
As a result, many in the automotive industry are hesitant
to adopt that technology for their future designs.  

The goal of this paper is to describe a simple architecture that has the following characteristics:
\begin{itemize}
\item the latency and switch memory occupancy have provable bounds that are easy to compute;
\item the switches have a simple design and are easy to configure;
\item the network integrates all the communications in the automobile with their specific latency and throughput requirements.
\end{itemize}

The key features of the architecture are as follows:
\begin{itemize}
\item each source bridge shapes and polices the traffic;
\item slow sources are aggregated on faster links;
\item the number of hops of fast flows is limited.
\end{itemize}

In this description, think of slow flows as being less than $2$Mb/s, as from devices attached to a CAN(-FD) bus, proximity sensors, push-button
switches, and the like.  Fast sources are the others: cameras, audio/video head units, radars, etc.

Thus, this architecture uses two central components of the IEEE time-sensitive networks standards: policing
and shaping.  It locates these components in the source bridges, instead of the network switches.
The main benefits of this approach are that the network uses a single priority class and that no further traffic reshaping
or scheduling is required in the switches.  As we explain in the paper, control signals are delivered with guaranteed
sub-millisecond latency even though they share queues with bursty best effort traffic and audio/video streams.  
These latency guarantees
are deterministic, not probabilistic: they are worst-case guarantees derived from analyzing the worst case
behavior of the network, not by simulations. The architecture ensures
that the web download of a new movie cannot interfere with the break signal from the self-driving control system
or from the break pedal.  Such guarantees are obviously necessary for the integration of the traffic on a single
network to be feasible. 

In this architecture, 
the configurations of the policer and shaper of a bridge are straightforward for, as we explain, they depend
only on the characteristics of the source attached to the bridge and not on the network topology, routing, and the other sources.
The configuration of the switches is straightforward
as they are only forwarding devices, with multicasting where necessary.  

The paper is organized as follows.  In section \ref{example} we start by describing a representative network that we use
to illustrate the methodology to calculate latency bounds.  Section \ref{analysis} explains the analytical results
behind these calculations.  Section \ref{analysisNetwork} presents the analysis of the representative network.  
Section \ref{generalization} discusses a free-rider principle which states that a fast network designed to transport
the signals from cameras and other high data rate sensors can transport the flows from CAN buses and other
slow sources for free. Section \ref{fast} explains a simple method to derive bounds on the latency and storage of fast flows.
Section \ref{conclusions}
summarizes the main points of the paper and gives hints on which future in vehicle network architectures may benefit from this work.

\section{Example} \label{example}

\begin{figure}[!h]
\centering \includegraphics[width=3.5in]{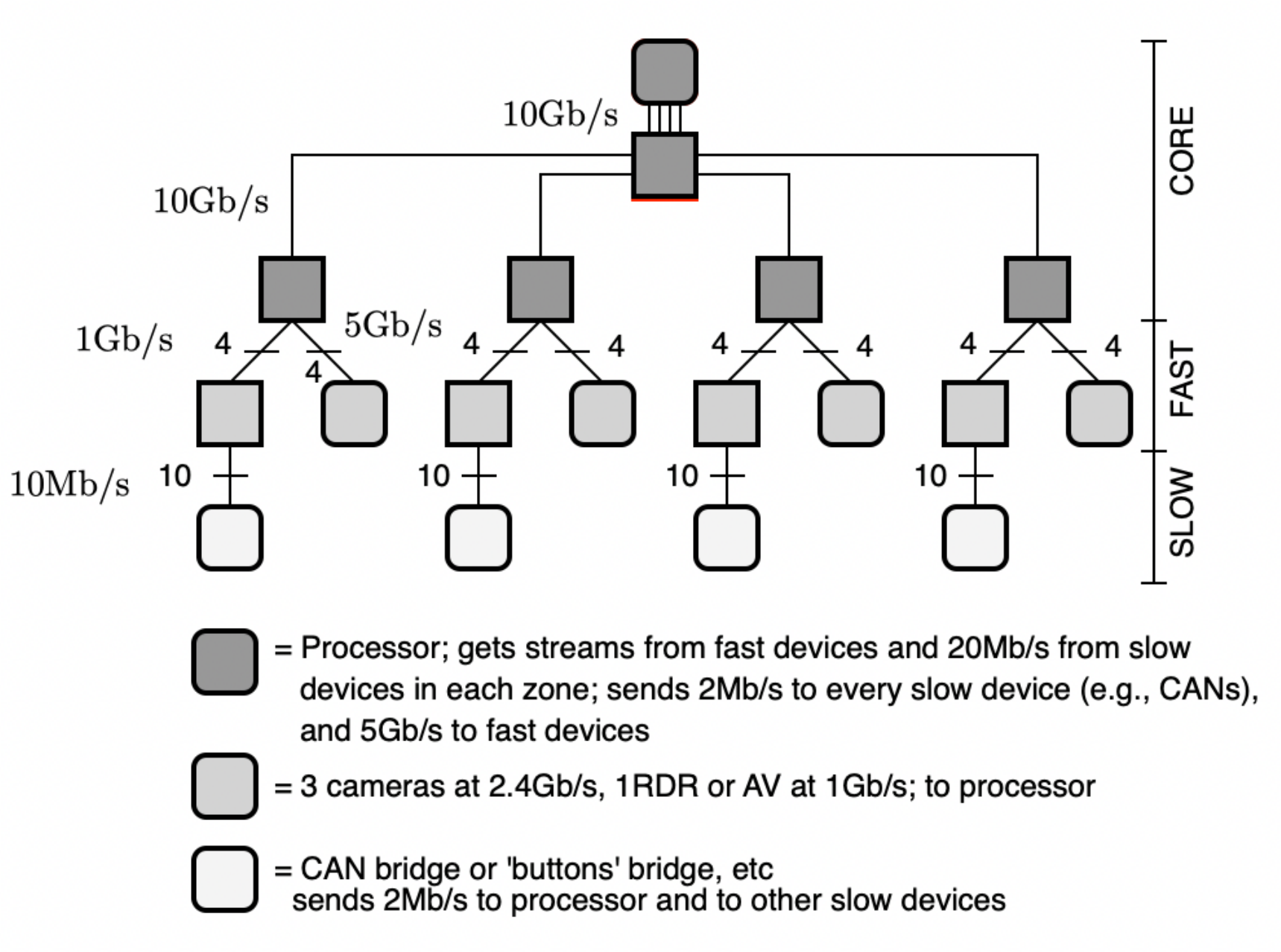} 
\caption{Network 1.}
\label{f1} 
\end{figure}

Figure \ref{f1} shows a representative network.  The network has three tiers of elements: core, fast, and slow
that correspond to the link rates indicated.
The squares with rounded corners
are devices that are sources and/or destinations of packets.  Possible elements are listed in the figure.  
Thus, ten slow devices are attached to a fast switch, each with a $10$Mb/s link; four fast switches are attached to 
a core switch, each with a $1$Gb/s link; four fast devices are also attached to a core switch, each with a $5$Gb/s link.

For instance, this network could have $12$ cameras, $12$ CAN devices bridges, $28$ button bridges, each connected to $20$ buttons, $1$ audio/video server, and three user interfaces. 

\subsection{Assumptions}

Each source, such as a camera, a radar, an audio/video server, a 5G antenna, a CAN electronic control unit, a button, and so on, is attached to a bridge that polices and shapes the traffic.  The policing verifies that the source is
not misbehaving, such as a faulty switch that thinks it is being pushed every millisecond.  The shaping separates the
packets by the maximum gap consistent with the deadline to deliver a group of packets or with the rate of a 
stream or file transfer.  For instance, say that a camera produces $2400$ packets of $1500$ Bytes every $16$ milliseconds and that one wishes to deliver these $2400$ packets in about $12$ milliseconds.  Then, the bridge
transmits one packet every $12$ms$/2400 = 5\mu$s. As another example, a bridge for a 5G antenna could send one $1500$ Byte sized packet every $10\mu$s 
to carry a data rate of $1.2$Gb/s.  Thus, every source of traffic is shaped, including best effort sources. 

We assume that three of the four fast devices attached to a core switch send $12,000$b (or $1,500$ Bytes) packets every $5\mu$s and one sends
$12,000$b packets every $15\mu$s.

The slow devices transmit packets of $512$b (or $64$ Bytes) that are separated by at least $250\mu$s.  
Thus, the rate of a slow device is bounded by $2$Mb/s.  
This spacing is compatible with CAN(-FD) like buses and
with push-button switches. For instance, the target end-to-end delay for engine control is $5$ms (see \cite{D07}).

For simplicity, we ignore switch processing delays and the minimum inter-frame gap.
That is, we model a switch port as a first-come, first-served queue that transmits a packet
of $B$ bits in $B/R$ seconds, where $R$ is the line rate in bits per second.

In the network of
Figure \ref{f1}, let us call a zone the set of devices attached to one of the four lower
core switches.  Thus, there are four zones.  We assume that at most $N$ flows
originate from each slow device attached to one $10$Mb/s link in one zone 
and that these flows carry packets of length
$512$b.  We also assume that at most $N$ such flows are destined to the slow
devices attached to a given $10$Mb/s of a given zone.  Thus, at most $10N$
such flows flow on a given $1$Gb/s link and at most $40N$ flow on a $10$Gb/s link.
We will assume $N = 4$ for this example.  Thus, up to four slow flows of $2$Mb/s share a $10$Mb/s link,
so that the utilization of that link is $80\%$. 

The spacing of the packets by a bridge creates gaps that limit the delays of other
packets.  To see this, consider a switch with $P$ input ports and assume that the incoming flow
on each port consists of packets of size $B$ bits separated by at least $T$ seconds.  Assume that
these flows go out on the same output port of the switch and that the output line rate is $R$ bits per second.
Observe that if $P\times B/R < T$, then the maximum delay of any packet through the switch is bounded by
$P\times B/R$ seconds.  Indeed, in the worst case, packets from the $P$ input ports arrive at the same time.
The last transmitted packet then has to wait for the transmission of the other $P - 1$ packets, and then for
its own transmission, before it has completely left the switch after $P\times B/R$ seconds. These $P$ packets are all transmitted
by time $T > P\times B/R$ before any other packet arrives.  Thus, even though the flows share the same queue and
the packets are transmitted in a first-come, first-served order, the delay of every flow is bounded.  

Note however that, as they leave the switch, the packets are no longer separated by a given gap.  Indeed,
they might leave the switch in bursts of $P$ packets that follow one another back-to-back.  One could
enforce a minimum gap between these packets by reshaping the output flow.  It turns out that such a 
reshaping is not necessary if the `burst accumulation' of the flows is bounded.  The analysis of the next sections
shows that this is the case in a network designed with some basic guidelines.

\section{Cruz Model} \label{analysis}

In \cite{cruz}, Rene Cruz defines a flow
as being of type $(A, a)$ if it can carry at most $A + at$ bits in any interval of $t$ seconds.
We call $A$ the burst size of the flow (in bits) and $a$ its sustained rate (in bits per second).
For instance, the flow from one of the cameras is of type $(A, a)$ where $A = 12,000$b and
$a = 2.4$Gb/s.  Also, the flow from a slow device is of type $(B, b)$ where $B$ is equal to $512$ bits
and $b \leq 2$Mb/s.  

From this definition, it follows that the superposition of a flow of type $(A, a)$ and one of type $(B, b)$
is of type $(A + B, a + b)$. Indeed, the superposition carries at most $A + B + (a + b)t$ bits in any interval of $t$ seconds.

Rene Cruz then analyzes a flow of type $(A, a)$ and one of type $(B, b)$ that share 
a first-come, first-served queue with service rate $R$ (in bits per second), where $R > a + b$. He observes the following properties:

\begin{lemma}[Cruz] \label{L1}

(a) As they leave the queue, 
the two flows are of type $(A', a)$
and $(B', b)$, respectively, where $A' \leq A + B(a/R)$ and $B' \leq B + A(b/R)$.  

(b) One has $A' \leq A + B$ and $B' \leq A + B$. 

(c) The queue stores at most $A + B$ bits.

(d) The delay through the queue is at most $(A + B)/R$ seconds. 

(e) The superposition of the two flows
as they exit the queue is of type $(C', c)$ with $c = a + b$ and $C' \leq A + B$.

(f) More generally, the burst size of a superposition 
of all the flows at the output of a queue is always
less than or equal to the sum of their burst sizes as they enter the queue.

\end{lemma}

\begin{proof}

(a) The maximum burst size of the first flow occurs when the $B$ bits arrive just before the 
$A$ bits, so that the queue first serves $B$ bits in $B/R$ seconds. During that time, $a(B/R)$
additional bits of the first flow enter the queue. Eventually, since $R > a + b$, at some time $t + B/R$,  the 
backlog of the queue gets transmitted and the queue catches up with the arrivals.  During $[B/R, t + B/R]$,
the queue transmits $A + a(B/R) + at$ bits of the first flow, which shows that it
is of type $(A + a(B/R), a)$.  A similar argument holds for the other flow.

(b) One has $A' \leq A + B(a/R) \leq A + B$ since $a < R$. Similarly, $B' \leq A + B$.

(c) The maximum storage in the queue occurs when the $A$ and $B$ bits
arrive at the same time.  

(d) The maximum delay in the queue occurs when the maximum storage is $A + B$.
In that case, it takes $(A + B)/R$ seconds to transmit the last one of these $A + B$ bits.

(e) To see why $C' \leq A + B$, consider some interval of duration $t$. The queue can serve the maximal
number of bits during that interval when it has its maximum occupancy $A + B$ at the
start of the interval. During the interval, the queue can serve at most $A + B$ plus the number $(a + b)t$
of bits that arrive during the interval.  

(f) This fact follows from the same argument as (e).
\end{proof}

The increase of the burst size of a flow from $A$ to $A'$ is precisely the burst accumulation that can occur in
a switch.  As we explained in the previous section, the goal is to characterize when this burst accumulation
is bounded by a value that guarantees that the delays through the switches are consistent with the target
end-to-end latency of the flows.

To simplify the algebra, we make the following observation:

\vskip 0.1in

\noindent
{\bf Small Flow Approximation.}  $B' = B + A(b/R)$ implies that $B' \approx B$ if $b/R \ll 1$.  This means that the
burst size of a flow with $b / R \ll 1$ does not increase much.  We say that such a flow is {\em small} and we call
the approximation $B' = B$ the {\em small flow approximation}.

We use this approximation to analyze Network 1 in the next section.

\section{Analysis of Network 1} \label{analysisNetwork}

\begin{figure}[!h]
\centering \includegraphics[width=3.5in]{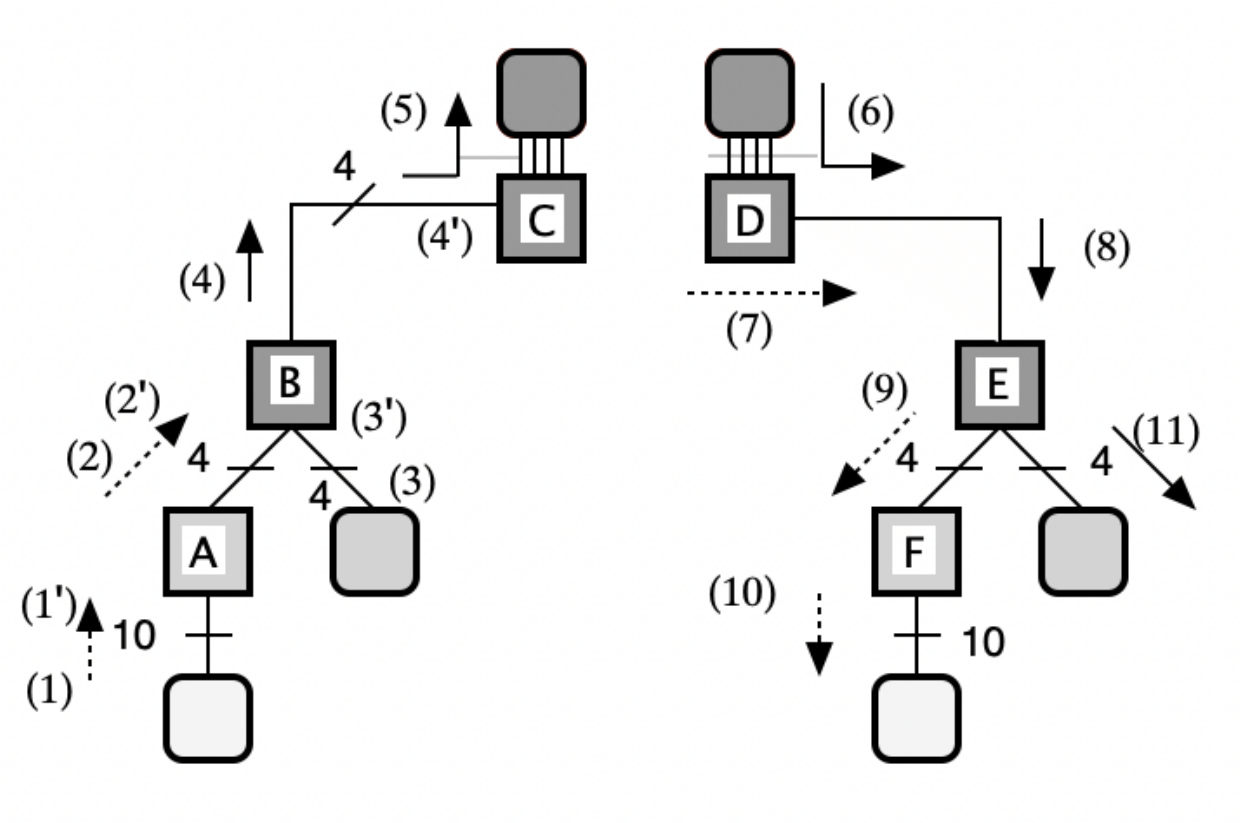} 
\caption{Links of network 1, where dotted arrows indicate slow flows.}
\label{f3} 
\end{figure}

For clarity, we
number the links of Network 1 as (1), ..., (11) as shown in Figure \ref{f3}.  The figure 
identifies the individual links and their
superposition (1'), ..., (4') as they arrive at a switch.  The switches are labelled A, ..., F.  The left-hand part
of the figure shows the flows going from the four zones to the central core switch and the processor
whereas the right-hand part shows the flows going from the central core switch and the processor
to a particular zone.  

Throughout this section, the quantity $A$ is equal to $12,000$ bits and $B$ is equal to $512$ bits.
(These quantities should not be confused with the labels A and B of the switches.) 

Let $f$ be a slow flow that originates in a slow device attached to link (1).  
Its burst
size is at most $B = 512$b, the rate of this flow is $2$Mb/s, so that the flow
is small on links (2), (4), (8), (9).  Using the small flow approximation, the burst size of
that flow remains very close to $B$ on those links.  

Since $N = 4$ slow flows share link (1), the delay on that link is bounded by
$(NB)/(10\mbox{Mb/s}) = 200\mu$s.

Consider the flows on link (9) that are destined to one of the links (10). By assumption,
there are at most $N = 4$ such slow flows. The burst size of that group of flows is bounded by
$N B = 2$kb. Accordingly, the delay on link (10) is bounded by $(2\mbox{kb})/(10\mbox{Mb/s}) = 200\mu$s.
Also, the memory occupancy on the southbound ports of switch F attached to slow devices is at most $10NB = 20$kb.
At most ten packets arrive at switch A from the links (1) and are transmitted on link (2)
as fast as they arrive. Thus, the delay on link (2) is bounded by $10B/(1\mbox{Gb/s}) = 5.1\mu$s.
Also, the storage on the northbound ports of switch A  is at most $10B = 5.1$kb.
Adding the occupancies of the southbound and northbound ports, we find that the memory occupancy of switches A and F is at most $25$kb.

Next, consider link (8).  There are at most $40N$ slow flows arriving at port (8)
of switch D.  In addition, there may be burst of $4A$ bits that arrive at that port from the processor.  Hence, the memory
occupancy of that port is at most $40NB + 4A = 130$kb.  
The packets that arrive at switch C via the links (4') get transmitted on the northbound 
links of that switch to the processor as soon as they arrive.  Thus, the memory occupancy of each
of the northbound links of switch C is bounded by $A$.  Summing the occupancy of all the ports, we
find that the total memory occupancy of switches C and D is bounded by 
$4 \times 130$kb $+ 4A  = 568$kb.

The total burst size of the flows that
arrive at switch B from southbound ports is bounded by
$40NB + 4A = 130$kb.  Correspondingly, the delay on link (4)
is bounded by \\
$(130\mbox{kb})/(10\mbox{Gb/s}) = 13\mu$s.   Also, the
storage in the northbound port (4) of switch B is bounded by $130$kb.

Similarly, the total burst size of the flows
that arrive at switch E via link (8) is also bounded by $40NB + 4A = 130$kb,
which is a bound on the storage in the southbound ports of switch E.

Summing these values, we find that the total memory occupancy in switches B and E
is bounded by $260$kb.  

Consider a flow from the processor and destined to a $5$Gb/s link (11).
The rate of that flow is at most $5$Gb/s and its burst size is at most $4A$.
Thus, the burst size of that flow on (8) is at most 
$4A + (40NB)(5/5) = 130$kb. Hence, the delay on link (11) is
bounded by $(130\mbox{kb})/(5\mbox{Gb/s}) = 26\mu$s.  

The burst size of the $N$ flows that arrive on (9) and are destined
to a $10$Mb/s link (10) is bounded by $NB = 2$kb. Accordingly, the
delay on link (10) is bounded by $(2\mbox{kb})/(10\mbox{Mb/s}) = 200\mu$s.

\begin{figure}[h]
\centering \includegraphics[width=3.5in]{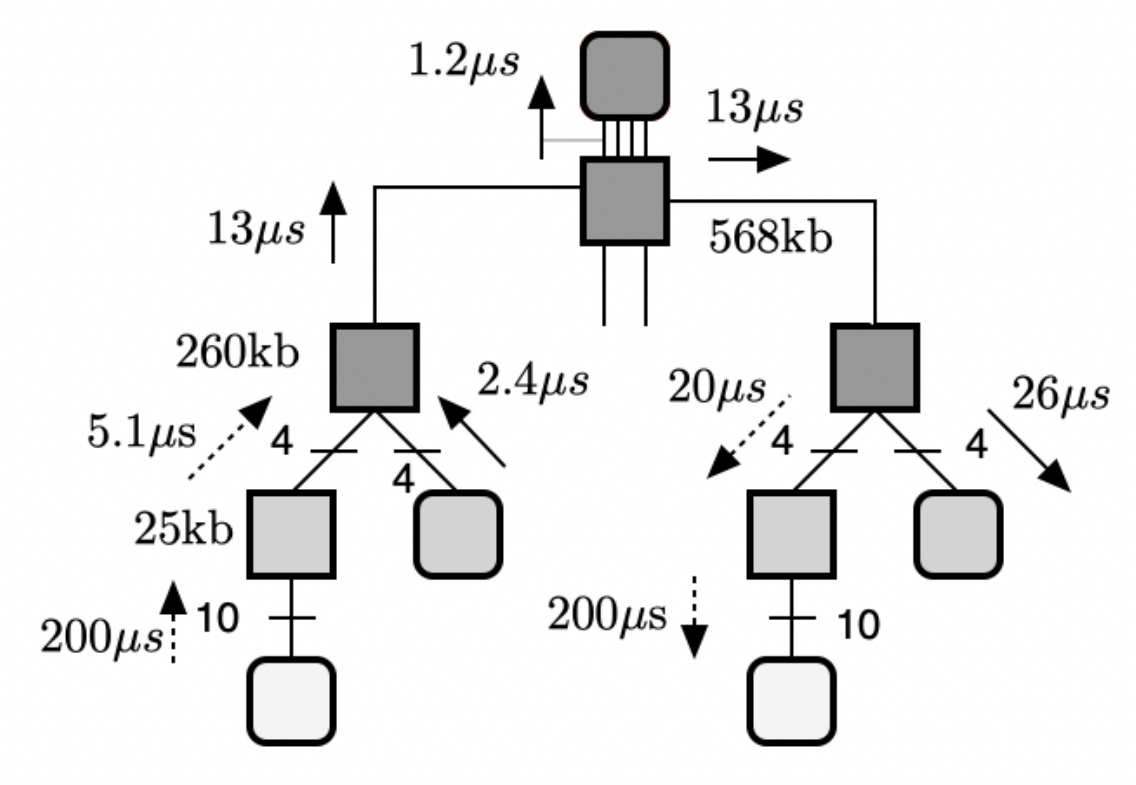} 
\caption{Delays and switch memory occupancy with the small flow approximation. The delay next to an arrow
is an upper bound on the latency through the output port of the switch attached to the corresponding link. The number
of bits next to a switch is the sum of the maximum occupancies of the output ports of the switch.}
\label{f4} 
\end{figure}

Figure \ref{f4} summarizes the results on delays and switch memory occupancy.

\begin{figure}[h]
\centering \includegraphics[width=3.5in]{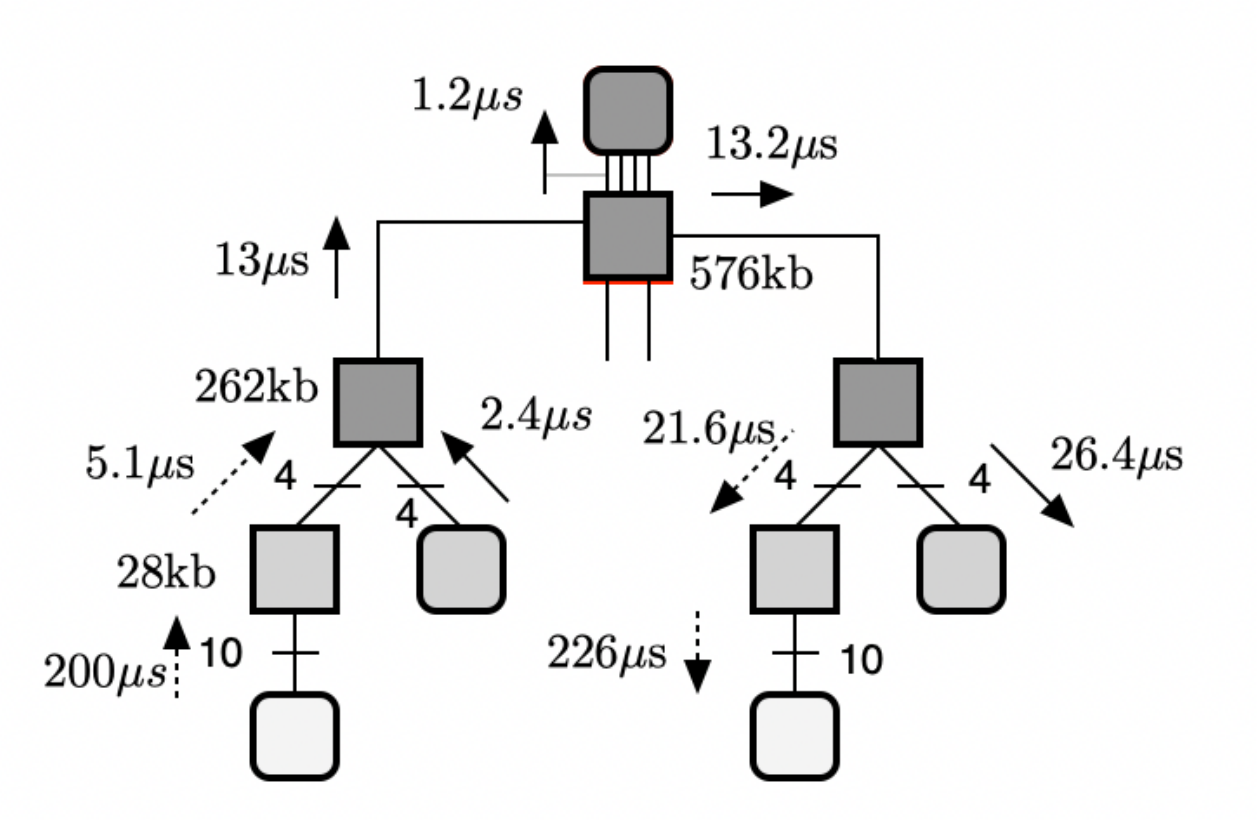} 
\caption{Delays and switch memory occupancy without the small flow approximation.}
\label{f4b} 
\end{figure}

If one uses the explicit formula $B' = B + A(b/R)$ for all the flows, instead of the small flow approximation
$B' = B$, the delay and memory occupancy values increase only by a few percent in this example.
Figure \ref{f4b} shows the values that one obtains by using the explicit formulas.
The details of the calculations are given in Appendix 1.

\newpage

\section{Free Rider Principle} \label{generalization}

In this section, we generalize the observations of the previous section.  
Our goal is to make precise the following intuitive statement:

\vskip 0.1in
\begin{center}
\noindent\fbox{%
    \parbox{2.6in}{%
        A fast network transports
slow flows for free.
    }%
}
\end{center}

For instance, a network designed to transport the data from cameras, radars,
lidars, audio/video servers to processors and user interfaces can also transport the slow flows
between CAN buses, push-buttons, relays, and so on.  

Let us clarify the meaning of this principle.  It is clear that if there is enough spare bandwidth on the
links of the fast network, then it can accommodate the rate of the slow flows. However, the ability
of the network to guarantee the required latency of the slow flows is much less obvious. It is precisely that
question that worries network designers when thinking of integrating flows from CAN buses with bursty
traffic from cameras, audio/video, and best effort flows.  Also, by `for free', one means without additional
reshaping, gating, and/or priority schedulers in the switches.

Thus, instead of having one network for cameras and for audio/video, and another to connect CAN buses,
one can use the first network to transport everything else.  Moreover, this integration 
requires only individual shaping of the sources by their bridges, as we already saw in
the case of Network 1.

To make the principle precise, we consider that the fast network has links with rates
$R_i$ for $i = 1, \ldots, I$.  For instance, in Network 1, the rates were $1$Gb/s, $5$Gb/s, $10$Gb/s.

Assume the following:
\begin{itemize}
\item At most $N_i$ slow flows go through a link with rate $R_i$, for $i = 1, \ldots, I$;
\item $R_i \gg b_i$ where $b_i$ is the maximum rate of a slow flow that flows through a link with rate $R_i$;
\item The maximum packet length of a slow flow is $B$;
\item The rate $R_i$ is large enough to accommodate the $N_i$ slow flows. That is, the spare bandwidth of link $i$
after taking the fast flows into account
exceeds $N_i b_i$.
\end{itemize}

Using the small flow approximation of section \ref{analysis}, we consider that the burst size of the slow flows
is approximately $B$ on every fast link.  Thus, the slow flows add a burst size $N_iB$ on a link
with rate $R_i$.  This increase in burst size has the following consequences:

\begin{enumerate}
\item the delay on a link with rate $R_i$ increases by at most $N_iB/R_i$ seconds;
\item the memory occupancy of a port with rate $R_i$ increases by at most $N_iB$ bits.
\end{enumerate}

To get concrete numbers, say that one can add $50$kb of memory storage per port and a delay of
$100\mu$s per port because of the slow flows.  Assume also that $B = 500$b. Then one can have
\[
N_i = 100
\]
provided that $R_i \geq 0.5$Gb/s and that the link has a spare bandwidth at least equal to $N_ib_i$ bits per second.

This implies that if the routing of the slow flows is such that at most $100$ slow flows share any given link
with the other flows, then the network carries them essentially for free.

\section{Designing the Fast Network} \label{fast}

Given the observations of the previous section, the main focus becomes the design of the fast network.
That is, we consider the network without the slow sources.  This network consists of the core switches and
the processors and fast devices attached to these core switches.  The goal of this section is
to explain how one can verify that this network achieves the desired latency objectives.

The burst size of a flow is at most $A$ when it leaves its bridge, where $A$ is the maximum packet size.  

\begin{figure}[!h]
\centering \includegraphics[width=3.5in]{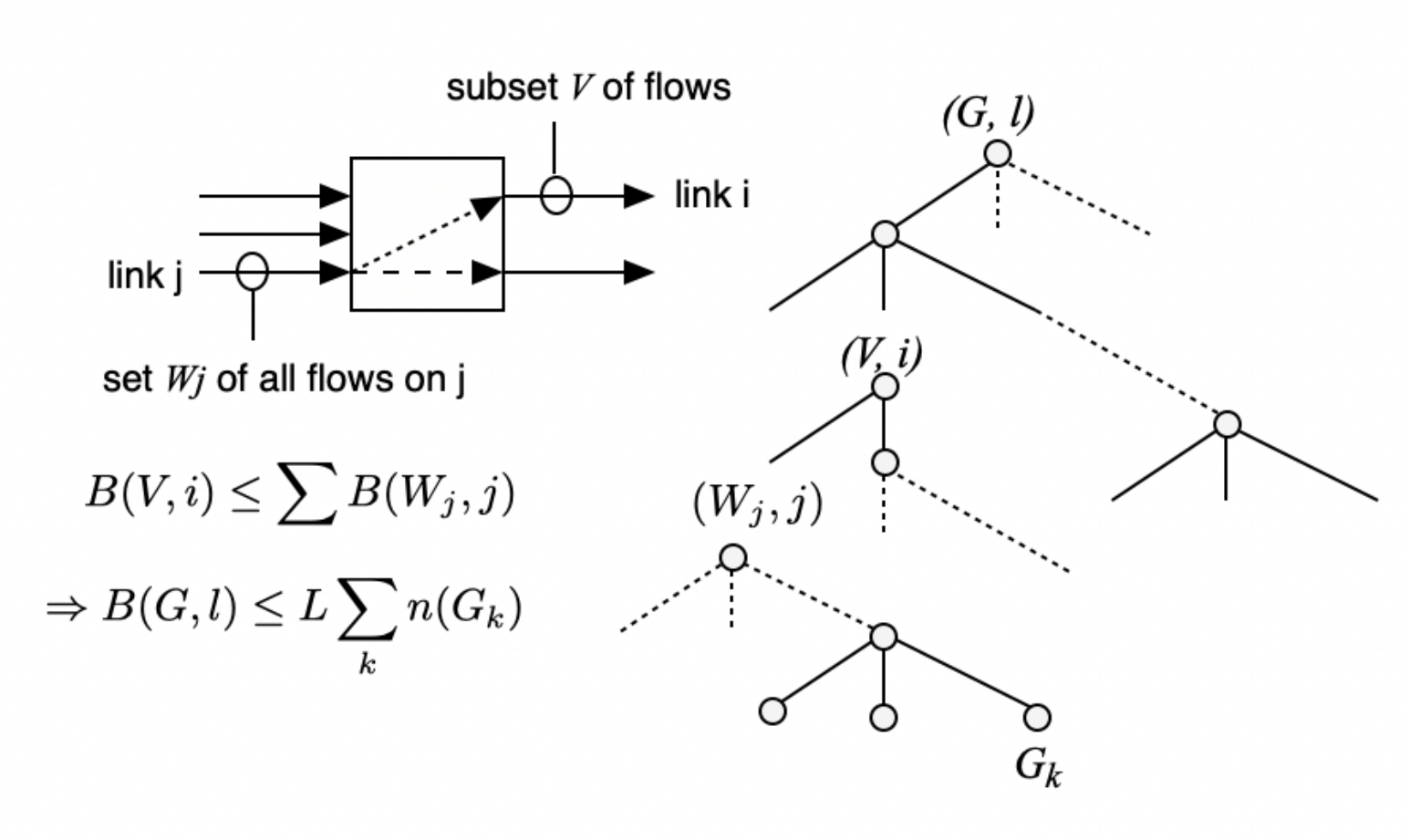} 
\caption{Tree for calculating an upper bound on the burst size $B(G,l)$ of a set $G$ of flows on a link $l$.}
\label{f5} 
\end{figure}

To calculate an upper bound on the burst size of a set $G$ of flows on a link $l$, one constructs a tree as follows.  (See Figure \ref{f5}.)
The root of the tree
is $(G, l)$. A node $(V, i)$ of the tree  is a set $V$
of flows on a link $i$. A child of $(V, i)$ is $(W_j, j)$ if $j$ is an input link of a switch with output link $i$
and $W_j$ is the set of flows on $j$  if at least one of the flows on $j$ uses link $i$. The tree has a finite depth.
The leaves of the tree are $\{(G_k, i_k), k = 1, \ldots, N\}$ where the links $i_k$ are attached to bridges.
Assume that the maximum packet size of each flow in each $G_k$ is $A$.  Then, the burst size of the set $G$ of flows
on link $l$ is bounded by $A \times \sum_k n(G_k)$ where $n(G_k)$ is the number of flows in $G_k$.

To see this, let $\beta(V, i)$ be the burst size of a set $V$ of flows on link $i$. Then, by Lemma \ref{L1}(f), one has $\beta(V, i) \leq \sum_j \beta(V_j, j)$ 
where $V_j$ is the subset of $V$ that uses link $j$.
Moreover, $\beta(V_j, j) \leq \beta(W_j, j)$ since $V_j \subset W_j$.  Hence, by induction on the tree, we find that
$\beta(G, l) \leq \sum_k \beta(G_k, l_k)$ where $l_k$ is the link attached to the bridge where $G_k$ originates.
Moreover, $\beta(G_k, l_k) \leq A n(G_k)$.

\begin{figure}[!h]
\centering \includegraphics[width=3.5in]{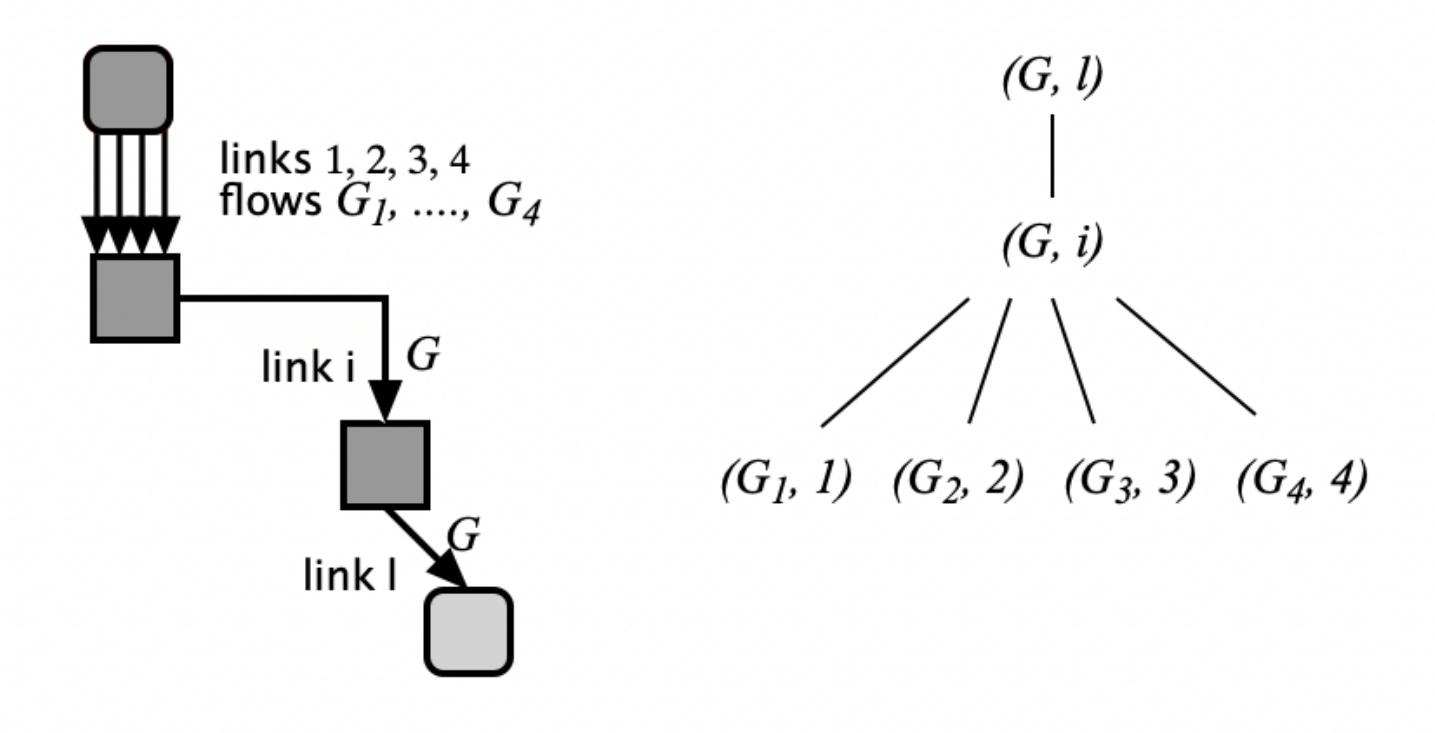} 
\caption{Tree for calculating an upper bound on the burst size $B(G,l)$ in the network of Figure \ref{f1}.}
\label{f6} 
\end{figure}

For instance, in the network of Figure \ref{f1}, consider the set $G$ of fast flows that are destined to a specific fast device
of the left-most zone and let $l$ be the link from the core switch to that fast device.  Figure \ref{f6} shows the tree rooted
at $(G, l)$.
In the figure, $G_1, \ldots, G_4$ consist of flows shaped by the processor, so that their burst size is bounded by $A = 12$kb.
Thus, the burst size $\beta(G,l)$ of $G$ on $l$ is bounded by $4A$.  

\section{Asymmetric Links}
Ethernet in general uses symmetric links, meaning that the line rates available in both directions are identical on any link segment. This has often been scrutinized for use-cases
like cameras, where the bandwidth required to transmit the video image is obviously much higher than the bandwidth required to control the camera. With the introduction of 
Energy Efficient Ethernet (EEE)\cite{IEEE-3az-EEE} this issue has partially been addressed. In EEE the power consumption of the lower bandwidth direction is reduced by turning off the transmitter while 
no data is available. Thus, the line rates at which the frames are transmitted remain symmetrical. As the Small Flow Approximation only depends on the line rate of a frame, it is completely 
untouched by EEE.

From a more academic point of view, one could also assume links to be asymmetric in line rate.
We examine the latency and memory utilization when links are designed to have a line rate equal to $125\%$ of the peak bandwidth of data required, independent from the standardized Ethernet modes of operation.  
That is, we design the links so that their maximum utilization is $80\%$.
We consider the case when the processor only sends slow traffic to fast devices (e.g., cameras), as it it the situation that corresponds to the slowest links which might result in larger latency and
memory occupancy.  In particular, we assume that the processor can send at most a burst of $4$ packets of $512$ bits to a given camera. The delay of these
four packets on the $10$Mb/s link attached to the camera is then at least $4 \times 512/(10\mbox{Mb/s}) = 200\mu$s.

\begin{figure}[!h]
\centering \includegraphics[width=3.5in]{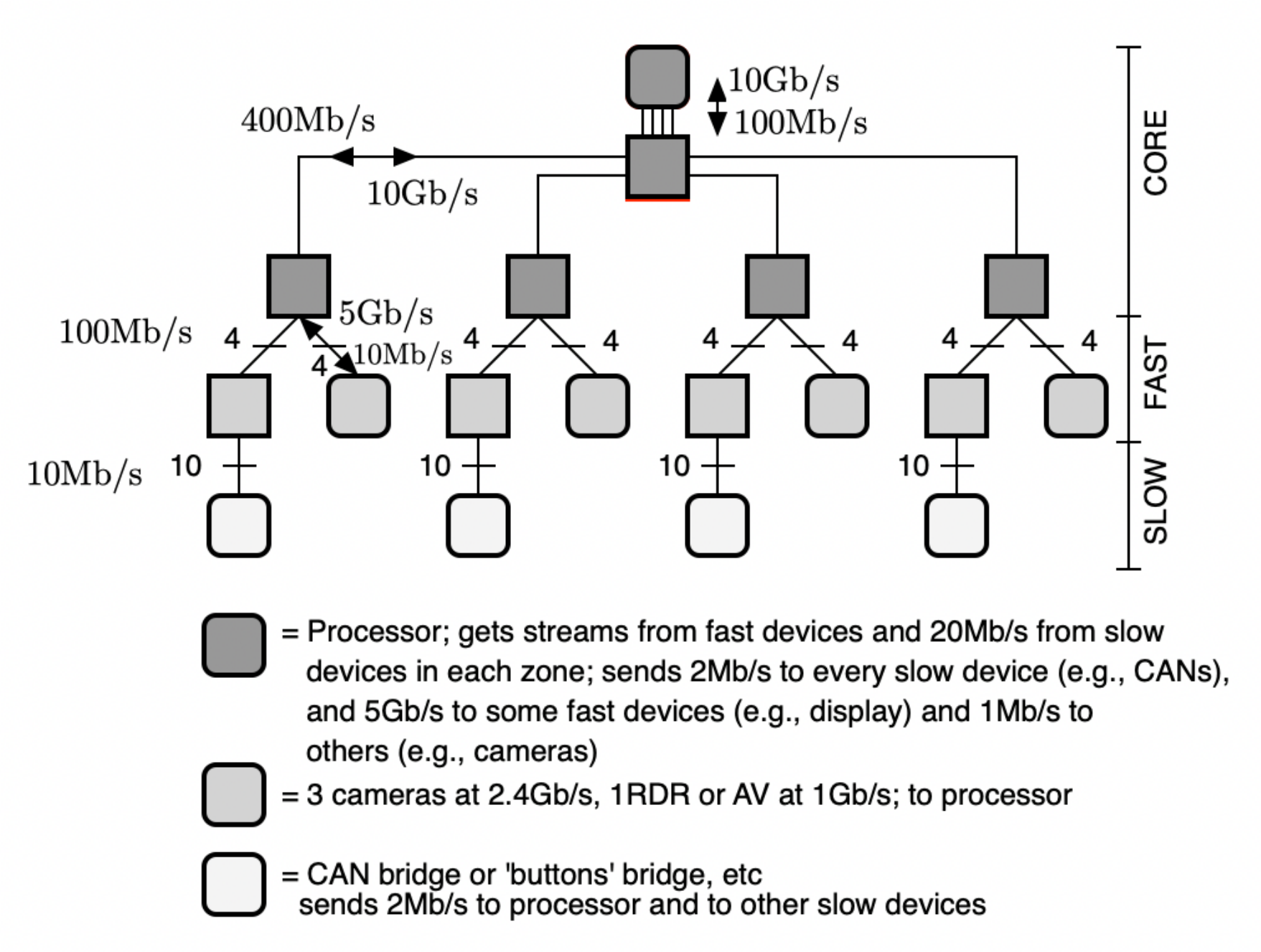} 
\caption{Network 2.}
\label{f10} 
\end{figure}

With the assumptions we made for the network traffic, the line rates are as shown in Figure \ref{f10}.

We analyze the latency and memory occupancy of this network as we did for the network of Figure \ref{f1}, without making the small flow approximation.  The details of the calculations are in 
Appendix 2 and the results are shown in Figure \ref{f11}.
\begin{figure}[!h]
\centering \includegraphics[width=3.5in]{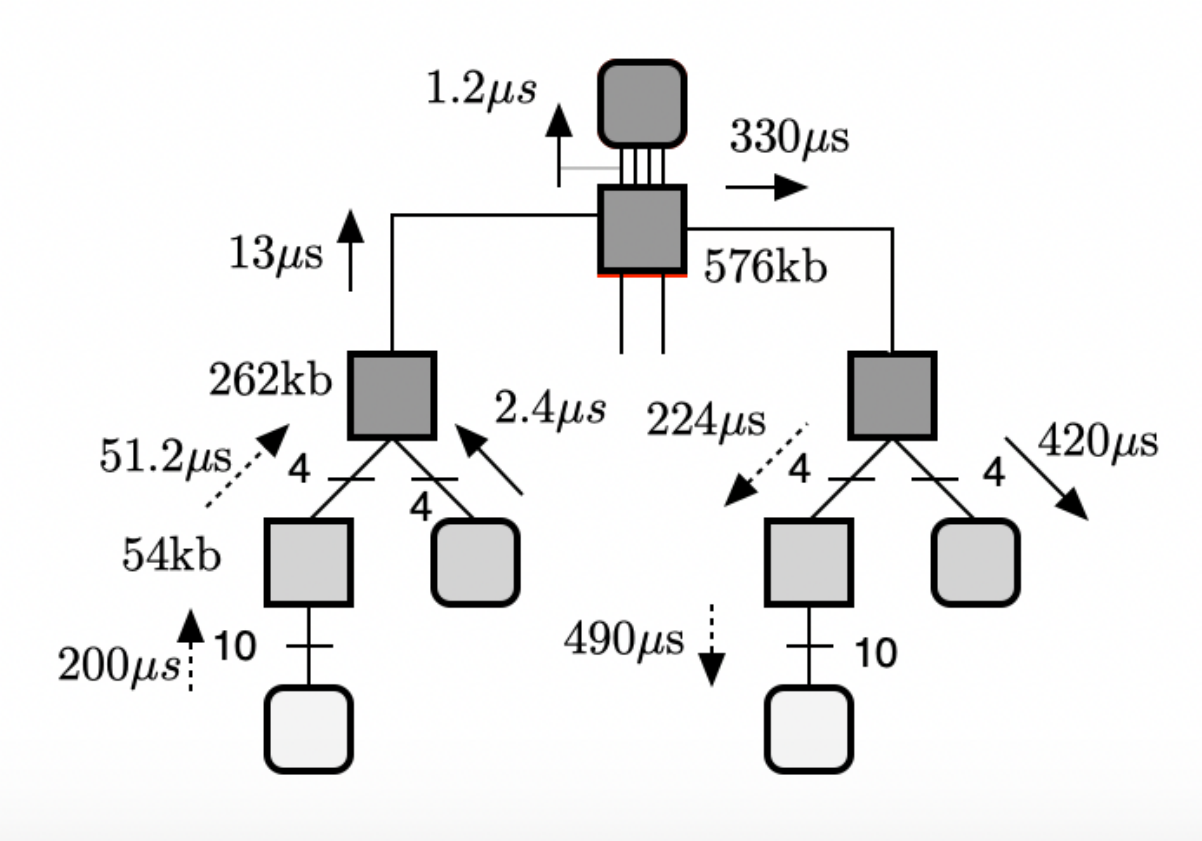} 
\caption{Delays and switch memory occupancy for Network 2.}
\label{f11} 
\end{figure}

For instance, the delay from a CAN bridge to another one in a different zone is bounded by $1.3$ms.
The bound was $0.5$ms with symmetric links (or with EEE links).

\section{Conclusions} \label{conclusions}

The paper discusses a simple network architecture for in-vehicle networks.  In this architecture, the bridge
attached to a source shapes its traffic by spacing out the packets as much as possible given the
deadline for transporting a burst of packets or given the sustained rate of the source.  The switches
use a single priority first-come, first-served queue and do not reshape traffic.  

The paper shows that this architecture can guarantee bounded latencies 
for all the flows and bounded memory storage for all the switches. 
Time-critical control messages are delivered with sub-millisecond latency.

The network devices are classified as core, fast, or slow.  Using 
Rene Cruz's results, one shows that a network fast enough to transport signals from cameras and
other fast sensors can also transport the data from slow sources and guarantee them their desired bounded latency.

One observation of the paper is that, because of the small flow approximation, one can focus on the design of the network that
transports the fast flows.  One can estimate the burst size of the fast flows by using 
a tree that captures the influence among flows.  

What are the prerequisites to allow the use of such a simple network architecture for in-vehicle networks? 

Due to the very different approaches of OEMs towards their product strategy and resulting network architectures, we need to classify them based on more abstract concepts in order to compare them. 
The most common starting point is and was a functional clustering. Here electronic control units (ECUs) which are generally related in their functionality and thereby often designed within the same organizational branch of the OEM \cite{ConwaysLaw} are connected together directly, offering only very limited interfaces to systems of other functional clusters. 
Such clusters may for example be the engine, the drive train, or the infotainment system. The theoretical extreme of this is often described as a domain-based system. 
How these different functional clusters or domains interact is very different for different OEMs. 
As the need to exchange data between domains has increased, e.g., to avoid duplication of expensive sensors in constrained packaging spaces, the so-called zonal model has gained much attention. 
In the zonal architecture the focus lies on integrating different functionalities onto a smaller set of ECUs, which share access to sensors and actuators.
This also leads to shorter cables and more importantly for this paper, a reduced number of hops in the network where particularly the aggregation of camera data requires high line rate links. 
Again the specific execution of this model could vary greatly between manufacturers and  few have actually come to the market so far. 
It could be argued that a shift away from privately owned individual vehicles towards automated robo-taxi fleets may drive this transition from functional clusters to geometrical zones, as the aspects of functional variance between customer chosen options, generational carry over and nameplate spread become less important for automated robo-taxi fleets while integration of systems and sensors becomes more important to solve the task of perception and control for driver-less operation in such vehicles. 
A further building block to allow this data convergence, which can be observed in the telecommunications business for quite some years already, is the only recent availability of multi-Gb/s Ethernet physical layer links for application inside a vehicle. All these aspects together create the environment wherein the concepts of this paper can be successfully deployed.

\section*{Acknowledgments}

The authors thank Georg Hoelger for his work on the analysis of shaping in TSN networks.
They also thank the team at Ethernovia for stimulating discussions.  That company designs
switches capable of implementing the IEEE TSN protocols and developed
software to evaluate the performance of networks and to optimize them. This paper is discussing only one of
many architectures possible with those switches.

\section*{Appendix 1}

This appendix provides the details of the calculations of the upper bounds on latency and memory
occupancy for network 1 in Figure \ref{f1} when one does not use the approximation $B' = B$ for slow flows.  These details show
why the approximation is very good.

Throughout the appendix, $A = 12$kb and $B = 512$b.

\subsubsection*{Output Port (9)}

Consider the set $G$ of flows that use a specific link (9) in Figure \ref{f3}.  That set consists
of up to $10N$ flows destined for up to $10$ small devices. The rate $b(G)$ of these flows
is bounded by $0.08$Gb/s. Let then $G_1, \ldots, G_4$ be the subset of $G$ that uses the $4$ different links (4).
Let also $G^c$ be the set of flows that share link (8) with $G$ but that are not in $G$ and
$G^c_1, \ldots, G^c_4$ the subsets of $G^c$ that use the $4$ different links (4). Also, let $n(G^c_i)$ be the number of
flows in $G^c_i$, for $i = 1, \ldots, 4$.

For an arbitrary set $V$ of flows on some link (j), we define $\beta(V, j)$ to be
the burst size of $V$ on link (j). Using Cruz's formulas, one finds
\begin{eqnarray*}
&&\beta(G,8) \leq \beta(G,4') + [\beta(G^c,4') + 4A]b(G)/(10\mbox{Gb/s}) \\
&&\beta(G^c, 4') = \sum_i \beta(G^c_i, 4) \\
&&\beta(G^c_i,4) \leq \beta(G^c_i,2') + [40B + 4A] b(G^c_i)/(10\mbox{Gb/s}) \\
&&\beta(G^c_i,2') \leq n(G^c_i)B\\
&& \beta(G,4') = \sum_i \beta(G_i,4) \\
&& \beta(G_i,4) \leq n(G_i)B + [40B + 4A]b(G_i)/(10\mbox{Gb/s}).
\end{eqnarray*}
Putting these relations together and using the fact that $\sum_i n(G^c_i) \leq 40,
\sum_i n(G_i) \leq 40, \sum_i b(G_i) \leq 80$Mb/s, and $\sum_i b(G^c_i) \leq 80$Mb/s,
one gets $\beta(G,8) \leq 21,581$b.

This bound on the burst size of $G$ on (8) corresponds to a bound on the latency through
port (9) that is equal to $(21,581$b$)/(1$Gb/s$)= 21.6\mu$s and a storage of $22$kb.

\subsubsection*{Output Port (10)}

To analyze port (10), let $H$ be the set of flows that go through that port. By
assumption, the rate $b(H)$ of those flows is bounded by $8$Mb/s.  
Let also $H_i$ be the subset of $H$ that uses a particular link (4), $H^c$ the set of
flows that share (4') with $H$ but are not in $H$, and the $H^c_i$ the subsets of $H^c$
that use the different link (4).  One has
\begin{eqnarray*}
&&\beta(H,9) \leq \beta(H,8) \leq \beta(H,4') \\
&&~~~~~~~ + [\beta(H^c,4') + 4A]b(H)/(10\text{Gb/s}) \\
&&\beta(H,4') = \sum_i \beta(H_i,4) \\
&&\beta(H_i,4) \leq n(H_i)B + [160B + 4A]b(H_i)/(10\text{Gb/s}) \\
&&\beta(H^c,4') = \sum_i \beta(H^c_i,4) \\
&&\beta(H^c_i,4) \leq n(H^c_i)B + [160B + 4A]b(H^c_i)/(10\text{Gb/s}).
\end{eqnarray*}
Using $n(H) \leq 4, n(H^c) \leq 160$, $b(H) \leq 8$Mb/s, and $b(H^c) \leq 320$Mb/s, one finds
$\beta(H,9) \leq 2,260$b.  These values correspond to a maximum storage equal to
$2,260$b for output port (10) and a maximum delay equal to $226\mu$s though that
port.

\subsubsection*{Output Port (11)}

The set $V$ of flows on (11) come from the central processor.  
Let $V^c$ be the flows that share (8) with $V$ but are not in $V$.
One has
\begin{eqnarray*}
&&\beta(V,8) \leq 4A + \beta(V^c,4')b(V)/(5\text{Gb/s}) \\
&&\beta(V^c,4') \leq 4\beta(G,4') 
\end{eqnarray*}
where $G$ is as in the analysis of output port (10).
Since the line rate or (11) is $5$Gb/s link, $b(G) \leq 5$Gb/s,
which gives $\beta(V,8) \leq 132$kb, which  corresponds to maximum latency
of $26.4 \mu$s through that port and a storage of $132$kb.

\subsubsection*{Storage in switch B}

Let $Q$ be the set of flows that arrive at switch E via (8).  One has
\begin{eqnarray*}
&& \beta(Q,8) \leq \beta(Q,4') + 4A \\
&& \beta(Q, 4') \leq 4\beta(G,4') .
\end{eqnarray*}
This gives $\beta(Q,8) \leq 132$kb, which
corresponds to a maximum storage of $132$kb in the output ports (9) and (11) of switch E.

The packets that arrive at switch B from (2') and (3') have a burst size bounded by $160B + 4A = 130$kb,
which is an upper bound on the storage in the output port (4) of switch B. 

Summing up, the total storage in switches B and E is bounded by $132\text{kb} + 130\text{kb} = 262\text{kb}$.

\subsubsection*{Output Port (8)}

The burst size of the flows that arrive at (8) is bounded by 
$4 \beta(G,4') + 4A = 132$kb. This corresponds to a maximum storage equal to
$132$kb for port (8) and a maximum delay equal to $13.2\mu$s through that port.

\subsubsection*{Storage in switch C}

The storage in switch C is bounded by four times the burst size into port (8) plus $4A$ (the burst size
into the ports going to the central processor). This adds up to $4 \times 132$kb + $4A = 576$kb.

\subsubsection*{Output Port (4)}

The burst size of the flows that arrive at (4) is bounded by $160B + 4A = 130$kb.
This corresponds to a maximum storage equal to $130$kb for port (4) and
a maximum delay equal to $13\mu$s through that port.

\subsubsection*{Output Port (2)}

Packets from (1') to (2) face a maximum delay of $10 \times 512$b$/(1$Gb/s$) = 5.1\mu$s.
The storage in that port is bounded by $5.1$kb.

\subsubsection*{Storage in switch A}

The storage in switch A is bounded by the storage in port (2), i.e., $5.1$kb, plus ten times
the storage in port (10), i.e., $10 \times 2.26$kb. The total is bounded by $28$kb.

\subsubsection*{Summary}

The maximum storage in each switch is bounded by the sum of the maximum storage
in the different output ports of the switch.  Combining the values derived above
yields Figure \ref{f4b}.

\section*{Appendix 2}

This appendix calculates the delays and memory occupancy bounds for Network 2  of Figure \ref{f10}.
Throughout the appendix, $A = 12$kb and $B = 512$b.

\subsubsection*{Output Port (9)}

Consider the set $G$ of flows that use a specific link (9) in Figure \ref{f3}.  That set consists
of up to $10N$ flows destined to up to $10$ small devices. The rate $b(G)$ of these flows
is bounded by $0.08$Gb/s. Let then $G_i$ be the subset of $G$ that uses the $4$ different links (4).
Let also $G^c$ be the set of flows that share link (8) with $G$ but that are not in $G$ and the
$G^c_i$ the subsets of $G^c$ that use the $4$ different links (4). Also, let $n(G^c_i)$ be the number of
flows in $G^c_i$.
Let $\beta(V,i)$ be the burst size of a set $V$ of flows on link (i). Using Cruz's formulas, one finds
\[
\beta(G,8) \leq \beta(G,4') + [\beta(G^c,4') + 4A]b(G)/(400\mbox{Mb/s}) 
\]
where $\beta(G,4')$ and $\beta(G^c,4')$ are as in Appendix 1. 
This gives $\beta(G,8) = 22.4$kb.
This bound on the burst size of $G$ on (8) corresponds to a bound on the latency through 
port (9) that is equal to $(22.4$kb$)/(100$Mb/s$)= 224\mu$s and a storage of $22.4$kb.

Note that in this calculation we assume that the processor might send large packets to a fast
device that would be attached to a fast link.

\subsubsection*{Output Port (10)}

To analyze port (10), let $H$ be the set of flows that go through that port. By
assumption, the rate $b(H)$ of those flows is bounded by $8$Mb/s.  
Let also  $H^c$ be the set of
flows that share (4') with $H$ but are not in $H$.  One has
\begin{eqnarray*}
&& \beta(H,9) \leq \beta(H,8) \leq \beta(H,4') \\
&&~~~~~~~ + [\beta(H^c,4') + 4A]b(H)/(400\text{Mb/s})
\end{eqnarray*}
where $\beta(H,4')$ and $\beta(H^c,4') $ are as in Appendix 1.
This gives
$\beta(H,9) \leq 4.9$kb.  These values correspond to a maximum storage equal to
$4.9$kb for output port (10) and a maximum delay equal to $490\mu$s though that
port.

\subsubsection*{Output Port (11)}

The set $V$ of flows on (11) come from the central processor.  
Let $V^c$ be the flows that share (8) with $V$ but are not in $V$.
One has
\begin{eqnarray*}
&&\beta(V,8) \leq 4B + \beta(V^c,4')b(V)/(400\text{Mb/s}) \\
&&\beta(V^c,4') \leq 4\beta(G,4') 
\end{eqnarray*}
where $G$ is as in the analysis of output port (10).
Since the line rate or (11) is $10$Mb/s link, $b(G) \leq 10$Mb/s,
which gives $\beta(V,8) \leq 4.2$kb, which  corresponds to maximum latency
of $420\mu$s through that port and a storage of $4.2$kb.

\subsubsection*{Storage in switch B}

Let $Q$ be the set of flows that arrive at switch E via (8).  One has
\begin{eqnarray*}
&& \beta(Q,8) \leq \beta(Q,4') + 4A \\
&& \beta(Q, 4') \leq 4\beta(G,4') .
\end{eqnarray*}
This gives $\beta(Q,8) \leq 132$kb, which
corresponds to a maximum storage of $132$kb in the output ports (9) and (11) of switch E.

The packets that arrive at switch B from (2') and (3') have a burst size bounded by $160B + 4A = 130$kb,
which is an upper bound on the storage in the output port (4) of switch B. 

Summing up, the total storage in switches B and E is bounded by $132\text{kb} + 130\text{kb} = 262\text{kb}$.

\subsubsection*{Output Port (8)}

The burst size of the flows that arrive at (8) is bounded by 
$4 \beta(G,4') + 4A = 132$kb. This corresponds to a maximum storage equal to
$132$kb for port (8) and a maximum delay equal to $132$kb$/(400$Mb/s$) = 330\mu$s through that port.

\subsubsection*{Storage in switch C}

The storage in switch C is bounded by four times the burst size into port (8) plus $4A$ (the burst size
into the ports going to the central processor). This adds up to $4 \times 132$kb + $4A = 576$kb.

\subsubsection*{Output Port (4)}

The burst size of the flows that arrive at (4) is bounded by $160B + 4A = 130$kb.
This corresponds to a maximum storage equal to $130$kb for port (4) and
a maximum delay equal to $13\mu$s through that port.

\subsubsection*{Output Port (2)}

Packets from (1') to (2)  face a maximum delay of $10 \times 512$b$/(100$Mb/s$)  = 51\mu$s.
The storage in that port is bounded by $5.1$kb.

\subsubsection*{Storage in switch A}

The storage in switch A is bounded by the storage in port (2), i.e., $5.1$kb, plus ten times
the storage in port (10), i.e., $10 \times 4.9$kb. The total is bounded by $54$kb.

\subsubsection*{Summary}

The maximum storage in each switch is bounded by the sum of the maximum storage
in the different output ports of the switch.  Combining the values derived above
yields Figure \ref{f11}.

\ifCLASSOPTIONcaptionsoff
  \newpage
\fi

%

\begin{IEEEbiography}[{\includegraphics[width=1in,height=1.25in,clip,keepaspectratio]{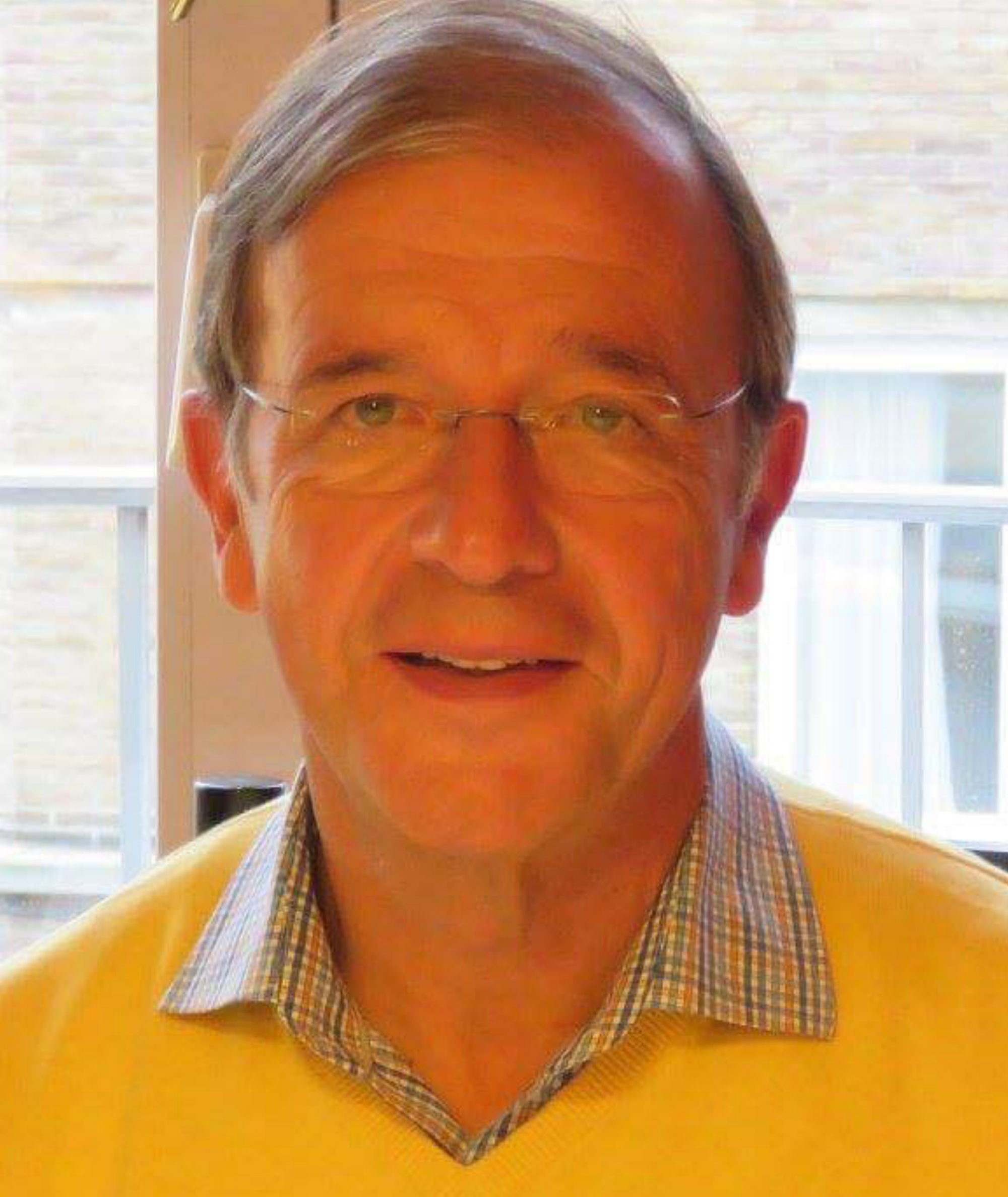}}]{Jean Walrand}
received his Ph.D. in EECS from UC Berkeley and has been on the faculty of that department since 1982. He is the author of An Introduction to Queueing Networks (Prentice Hall, 1988), Communication Networks: A First Course (2nd ed. McGraw-Hill, 1998), Probability in Electrical Engineering and Computer Science (2nd ed. Springer, 2021), and Uncertainty: A User Guide (Amazon, 2019) and co-author of High-Performance Communication Networks (2nd ed, Morgan Kaufman, 2000), 
Communication Networks: A Concise Introduction (2nd ed. Morgan \& Claypool, 2018), Scheduling and Congestion Control for
Communication and Processing networks (Morgan \& Claypool, 2010), and Sharing Network Resources (Morgan \& Claypool, 2014). 

His research interests include stochastic processes, queuing theory, communication networks, game theory and the economics of the Internet. 

Prof. Walrand is a Fellow of the Belgian American Education Foundation and a Life Fellow of the IEEE and a recipient of the Lanchester Prize, the Stephen O. Rice Prize, the IEEE Kobayashi Award and the ACM Sigmetrics Achievement Award.

\end{IEEEbiography}

\begin{IEEEbiography}[{\includegraphics[width=1in,height=1.25in,clip,keepaspectratio]{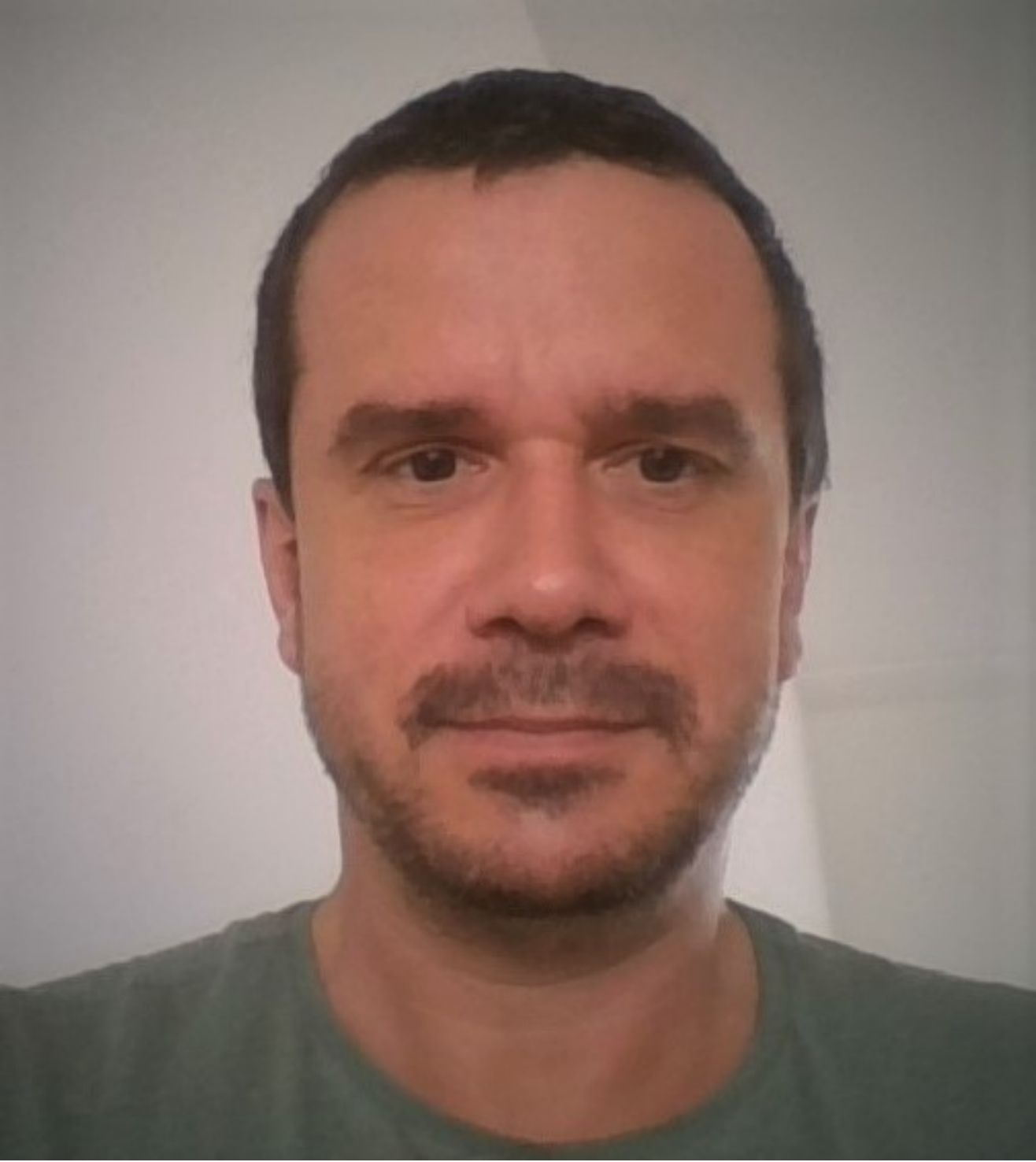}}]{Max Turner}
received his Dipl. Phys. from the Universit\"at Ulm, Germany in 1999. He joined BMW late in 2002 where he initially worked on MOST and FlexRay. 
During a stay in the USA he worked on V2x wireless systems and the DSRC standardization (IEEE802.11p).
Returning to Munich in 2008 Max worked on the introduction of Ethernet in Autosar and became part of the group creating the ISO 13400 `Diagnostics over IP' standard. 
For the following 10 years Max was a member of the team introducing Ethernet as a system-bus (including SOME/IP, XCP, DLT, AVB and other protocols) into all BMW vehicle generations.
For not quite two years Max joined Jaguar Land Rover in in the UK, where he gathered experience in the overall E/E architecture for automated vehicles as the lead architect.
Since Dec. 2019 Max serves as the automotive network architect for Ethernovia. He is and has been an active contributor to the AVB and TSN working groups of IEEE, OpenAlliance and AVnu for most of the time.
\end{IEEEbiography}

\begin{IEEEbiography}[{\includegraphics[width=1in,height=1.25in,clip,keepaspectratio]{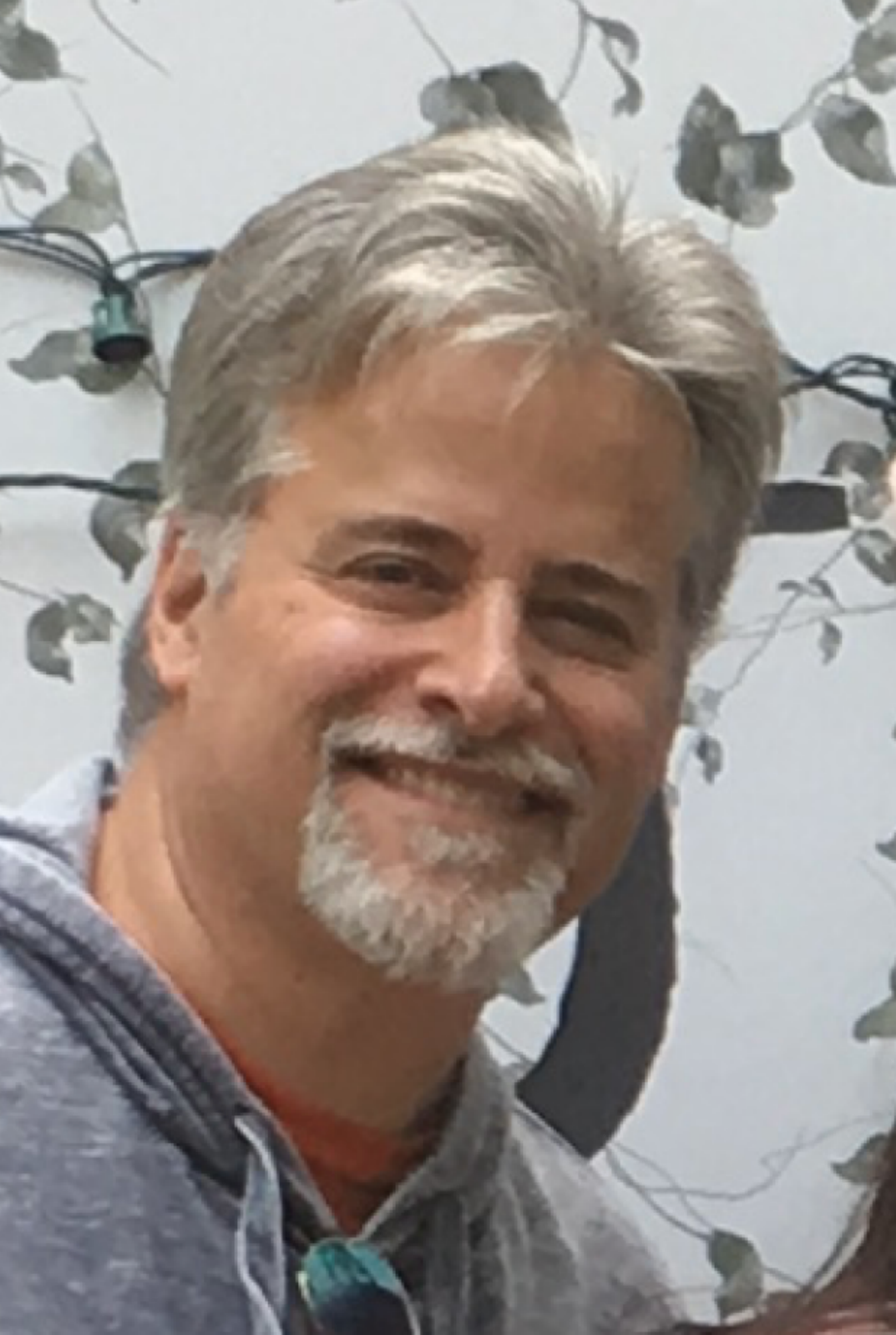}}]{Roy Myers}
received his Bachelor’s and Master of Science in Electrical Engineering from Georgia Institute of Technology in 1990 and 1991, respectively. He then joined National Semiconductor LAN Division designing mixed signal Ethernet IC products. In 1996, he joined Enable Semiconductor, as a founding member, developing Ethernet 100Base-TX transceivers. Enable Semiconductor was acquired by Lucent Technologies in March 1999. After leaving Lucent, he co-founded Terablaze Inc in 2000 where he was Director of Engineering developing highly scalable network fabrics and layer 2/3 Gigabit Ethernet switching solutions. TeraBlaze Inc was acquired by Agere Systems in 2004. Roy continued to lead the Gigabit Ethernet switch development at Agere and then at LSI after acquisition of Agere Systems in 2007.  He left LSI in 2007 to join Aquantia Corp (AQ:NYSE) where he was Chief Architect developing a series of 10GBase-T and Multi-Gig MAC/PHY products to serve client, enterprise, and data center markets. Since April 2018, Roy is a co-founder and SVP of Engineering of Ethernovia Inc developing network solutions for the next generation of automobiles. Roy holds 14 granted patents in communication technology.

\end{IEEEbiography}

\end{document}